\documentclass[amsmath,nofootinbib,twocolumn]{revtex4-1} 
\usepackage{aas_macros,cases,graphicx,subfigure}
\usepackage[colorlinks]{hyperref}
\pdfoutput=1

\newcommand{\ab}[1]{\left|#1\right|}
\newcommand{\br}[1]{\left[#1\right]}
\newcommand{\cu}[1]{\left\{#1\right\}}
\newcommand{\pa}[1]{\left(#1\right)}
\newcommand{\ed}{\,\mathrm{d}}
\newcommand{\pd}{\,\partial}
\newcommand{\C}{\mathcal{C}}
\newcommand{\I}{\mathcal{I}}

\begin{document}

\title{Pulsar Magnetospheres: Beyond the Flat Spacetime Dipole}

\author{Samuel E. Gralla} 
\affiliation{Department of Physics, University of Arizona, Tucson, AZ 85721, USA}
\author{Alexandru Lupsasca}
\affiliation{Center for the Fundamental Laws of Nature, Harvard University, Cambridge, MA 02138, USA}
\author{Alexander Philippov}
\affiliation{Department of Astrophysical Sciences, Princeton University, Princeton, NJ 08540, USA}

\begin{abstract}
Most studies of the pulsar magnetosphere have assumed a pure magnetic dipole in flat spacetime.  However, recent work suggests that the effects of general relativity are in fact of vital importance and that realistic pulsar magnetic fields will have a significant nondipolar component.  We introduce a general analytical method for studying the axisymmetric force-free magnetosphere of a slowly rotating star of {\it arbitrary} magnetic field, mass, radius, and moment of inertia, including all the effects of general relativity.  We confirm that spacelike current is generically present in the polar caps (suggesting a pair production region), irrespective of the stellar magnetic field.  We show that general relativity introduces a $\sim60\%$ correction to the formula for the dipolar component of the surface magnetic field inferred from spindown.  Finally, we show that the location and shape of the polar caps can be modified dramatically by even modestly strong higher moments.  This can affect emission processes occurring near the star and may help explain the modified beam characteristics of millisecond pulsars.
\end{abstract}

\maketitle

\section{Introduction}

Pulsars display a striking and diverse observational phenomenology, including pulsed emission across a variety of wavelengths and synchrotron emission in associated pulsar wind nebulae.  The challenge to the pulsar theorist is to explain this array of observations from the rather simple theoretical beginnings of a rotating, magnetized, neutron star.  After more than 40 years of effort, much has been understood, but the full solution remains elusive \cite{Arons2012}.

The presence of plasma outside the star is likely a key ingredient.  Early on, it was realized that the rotation of the star in vacuum produces enormous ``accelerating'' electric fields (i.e., fields with component along the magnetic field) outside the star \cite{Goldreich69}.  Unless the work function is unnaturally large, these fields lift charged particles off the star and accelerate them to ultra-relativistic energies.  The emitted curvature photons can then interact with the magnetic field (or other charged particles) to produce particle pairs \cite{Sturrock71}, which act to screen the accelerating field.  If this mechanism continuously produces charges to replace those that are accelerated away, one can hope to explain both the pulsed radio emission (from collective behavior in the bulk plasma) as well as the high energy particles in the wind.  To fill in the details, however, one needs (at a minimum) the magnetic geometry, which must be determined self-consistently by solving for the global structure of the plasma magnetosphere.

Considerable progress has been made in constructing global models of pulsar magnetospheres under the assumption of abundant plasma supply.  Since the energy of the magnetospheric plasma is much smaller than the energy in the magnetic field, the magnetohydrodynamic equations can be simplified to the low-inertia, or force-free, limit.  For example, self-consistent force-free numerical solutions of axisymmetric \cite{Contopoulos99, Gruzinov05, McKinney06, Timokhin06} and oblique \cite{Spitkovsky06, Kalapotharakos09, Li11, Kalapotharakos12, Petri12} pulsar magnetospheres were developed. These studies were extended to include the effects of plasma pressure and inertia \cite{Komissarov06, Tchekhovskoy13}, as well as general relativity \cite{Lehner12, Palenzuela13, Ruiz14, Petri16}, which are important in current layers and near the star, respectively.  This has produced a canonical model for the dipole pulsar magnetosphere, with polar outflows along open field lines separated from a zone of closed field lines by thin return current layers, together with a strong current sheet beyond the light cylinder.  Despite this success, the basic questions remain: is the plasma really there?  And can it produce pulsed emission and relativistic particles for the wind?

Addressing these questions from first principles requires a full kinetic treatment, including the physics of pair formation and particle acceleration.  Recent years have seen significant progress in kinetic plasma simulations, which have shed light on how and when plasma can be produced in a magnetosphere.  Working with a dipolar field in flat spacetime, Refs.~\cite{Chen14} and \cite{Philippov15a} found that, provided a source of particles at the stellar surface and allowing pair production close to the light cylinder, a nearly force-free magnetosphere develops in which pairs are produced in the thin current layers and the equatorial current sheet.  Notably, however, there is no pair formation in the bulk of the polar cap in these models.  Instead, the accelerating electric field is screened by a mildly relativistic charge-separated flow of particles lifted from the neutron star surface \cite{Shibata97, Beloborodov08, Timokhin13}.

This slow outflow does not radiate the gamma photons needed to produce pairs, and is not promising for producing radio emission or pulsar wind particles.  However, Ref.~\cite{Philippov15b} showed that the inclusion of general relativistic (GR) effects, in particular the dragging of inertial frames around the neutron star, leads to the ignition of a discharge at the polar cap of the aligned rotator.  This phenomenon is associated with the current density becoming larger than the charge density (times the speed of light), which can no longer be supported by one sign of escaping charge and requires pair production.  In relativistic language, the charge-current four-vector is \textit{spacelike}.  Since the configuration is force-free, this suggests a simple prescription for studying magnetospheric discharges \cite{Timokhin13}: solve the equations of force-free electrodynamics and find the regions near the star where the current is spacelike.

Reference~\cite{Philippov15b} studied the aligned dipole in GR and found qualitative agreement between the spacelike-current regions of a force-free simulation and the pair-formation regions in the kinetic simulation.  This supports the method, but much remains to be explored.  In particular, there is little reason to believe that real pulsars have pure dipolar magnetic fields.  Indeed, the magnetic field is determined by complicated internal processes, and to the extent that these are understood, the field is not dipolar \cite{Geppert14, Gourgouliatos16}.  In this paper, we introduce an analytical procedure to determine the near-field magnetosphere, including the globally determined current flow, of a slowly rotating star with an \textit{arbitrary} axisymmetric magnetic field profile.  We use the method of matched asymptotic expansions, solving the equations of force-free electrodynamics on curved spacetime to linear order in $\epsilon=R_\star/R_L$, the ratio of the stellar radius to the light cylinder radius.

We show that spacelike current generically appears in the current flow regions near the star, or ``polar caps,'' thereby extending the dipolar results of Ref.~\cite{Philippov15b} to generic fields.  We show, however, that the location and shape of the polar caps changes significantly when higher multipoles are included.  In particular, a polar cap can occupy a thin annular region of angular width $\sim\epsilon$ displaced from the rotational pole instead of the classic $\sim\sqrt{\epsilon}$ region surrounding the pole.  This could give rise to double-peaked emission and/or help explain the discrepant beam size scalings between ordinary and millisecond pulsars \cite{Kramer98}.  Such shifted polar caps are also further favorable for pair production because of higher field line curvature and subsequent increase of the magnetic opacity \cite{Barnard82}.  In general, our analytical method provides the detailed near-field magnetosphere and may hence be used to construct models for near-zone pair creation and emission based on more complicated and realistic magnetic fields \cite{Arons2012}.

We also consider the spindown luminosity of the magnetosphere.  We show that the luminosity depends only on the dipolar component of the surface magnetic field.  We find that general relativity significantly decreases the luminosity for a fixed magnitude of the surface magnetic field, corresponding to a $\sim60\%$ correction to the formula for the surface magnetic field inferred from measurements of the pulsar period and period derivative.  This correction is distinct from that found recently in Refs.~\cite{Ruiz14, Petri16}, which occurs only for rapid spin.  In these works, the comparison to flat spacetime was made at fixed asymptotic dipole moment.  Defined this way, there are no corrections due to general relativity for slowly spinning pulsars.  In general, simulations are limited to larger values of $\epsilon=R_\star/R_L\sim0.2$, whereas our analytical method explores the complimentary regime $\epsilon\to0$, which is valid for most pulsars.

The paper is organized as follows.  In Sec.~\ref{sec:GeneralMethod}, we describe our matched asymptotic expansion method and derive the spindown luminosity as well as general formulas for the polar cap regions and the near-zone current.  We then apply the method to pulsars with dipole fields (Sec.~\ref{sec:Dipole}), both dipole and quadrupole fields (Sec.~\ref{sec:Quadrudipole}), and more general configurations suggested by recent work on neutron star magnetic fields (Sec.~\ref{sec:Models}).  In Sec.~\ref{sec:Discussion}, we summarize and discuss observational implications.  We use Heaviside-Lorentz units and set $G=c=1$.

\section{General Method}
\label{sec:GeneralMethod}

We consider a slowly rotating star in general relativity.  To linear order in the rotation velocity, the star is spherical and the metric outside is given by \cite{Thorne68}
\begin{align}
	\label{eq:StarMetric}
	ds^2&=\alpha^2\ed t^2+\alpha^{-2}\ed r^2\\
	&\quad+r^2\br{\!\ed\theta^2+\sin^2{\theta}\pa{\!\ed\phi-\Omega_Z\ed t}^2},
	\quad r>R_\star,
	\nonumber
\end{align}
where the ``redshift factor'' $\alpha$ and the ``frame-drag frequency'' $\Omega_Z$ are
\begin{align}
	\label{eq:AlphaOmegaDefinition}
	\alpha&=\sqrt{1-\frac{2M}{r}},\qquad\Omega_Z=\frac{2\hat{I}}{r^3}\Omega.
\end{align}
Here, $R_\star$ is the (areal) radius, $M$ is the mass, $\Omega$ is the angular velocity, and $\hat{I}$ is the moment of inertia (defined as the angular momentum over the angular velocity).  The angular velocity defines the ``light cylinder'' on which a co-rotating observer would move with the speed of light.  We define
\begin{align}
	R_L=\frac{1}{\Omega},
\end{align}
which agrees with the actual light cylinder radius in the slow-rotation limit.

We can characterize the problem by an overall scale and three dimensionless parameters,
\begin{align}
	\epsilon=\frac{R_\star}{R_L},\qquad\C=\frac{2M}{R_\star},\qquad\I=\frac{\hat{I}}{MR_\star^2},
\end{align}
corresponding to the surface rotation velocity, the stellar compactness, and the dimensionless moment of inertia, respectively.  The compactness is subject to the theoretical bound $\C<8/9$ \cite{Buchdahl59}; realistic neutron star models generally have $\C\sim1/2$ \cite{Haensel07}.  Note that on the surface of the star, we have
\begin{align}
	\label{eq:AlphaOmega}
	\alpha=\sqrt{1-\C},\qquad\Omega_Z=\C\I\Omega.
\end{align}
The metric \eqref{eq:StarMetric} is valid to linear order in $\epsilon$.  In the remainder of the paper, we will compute physical quantities to leading nontrivial order in $\epsilon$.\footnote{In perturbation theory, leading nontrivial order (or ``leading order'') means the first order at which a given quantity is nonzero.}

If the star is surrounded by a stationary, axisymmetric, force-free magnetosphere, then the electromagnetic field is characterized by the magnetic flux function $\psi(r,\theta)$, the field line rotation velocity $\Omega_F(\psi)$, and the polar current $I(\psi)$ as \cite{Gralla14}
\begin{align}
	\label{eq:F}
	F=\frac{I(\psi)}{2\pi\alpha^2\sin{\theta}}\ed r\wedge\!\ed\theta+\!\ed\psi\wedge\br{\!\ed\phi-\Omega_F(\psi)\ed t}.
\end{align}
We use the conventions of Refs.~\cite{Tchekhovskoy10,Gralla15}.  For the field configuration to be smooth, the flux function and polar current must vanish quadratically in $\theta$ at both poles.  Level sets of the flux function are called poloidal field lines, or field lines for short.

We assume that the star is perfectly conducting, which fixes the field line rotation to the stellar rotation,
\begin{align}
	\label{eq:OmegaField}
	\Omega_F(\psi)=\Omega.
\end{align}
The force-free condition implies that $\psi$ and $I(\psi)$ satisfy
\begin{align}
	\label{eq:StreamEquation}
	r^2\pd_r\!\pa{\alpha^2\chi\pd_r\psi}&+\sin{\theta}\pd_\theta\!\pa{\frac{\chi}{\sin{\theta}}\pd_\theta\psi}\\
	=\,&-\pa{\frac{r}{2\pi\alpha}}^2I(\psi)I'(\psi),
	\nonumber
\end{align}
where we introduced
\begin{align}
	\chi=1-\pa{\frac{r\sin{\theta}}{\alpha}}^2\pa{\Omega-\Omega_Z}^2.
\end{align}
The star's magnetic field provides a boundary condition
\begin{align}
	\label{eq:NearFluxBoundaryCondition}
	\psi\big|_{r=R_\star}=\psi_\star(\theta),
\end{align}
where $\psi_\star(\theta)$ is the flux function on the stellar surface.  Since the pulsar is isolated, the field lines should become radial at large $r$, i.e.,
\begin{align}
	\label{eq:FarFluxBoundaryCondition}
	\psi=A(\theta)+\mathcal{O}\!\pa{r^{-1}},
\end{align}
for some $A(\theta)$ to be determined by solving the equations.  Together with the stream equation \eqref{eq:StreamEquation}, this condition implies
\begin{align}
	\label{eq:FarCurrentBoundaryCondition}
	I=2\pi\Omega\sin{\theta}\pd_\theta\psi,\quad r\to\infty,
\end{align}
which is equivalent to having electric and magnetic fields of asymptotically equal magnitude as $r\to\infty$.

This defines the problem: Given a choice of stellar parameters $\cu{R_\star,M,\hat{I},\Omega,\psi_\star(\theta)}$, one must solve Eq.~\eqref{eq:StreamEquation} for $\psi$ and $I(\psi)$ satisfying the boundary conditions \eqref{eq:NearFluxBoundaryCondition} and \eqref{eq:FarCurrentBoundaryCondition}.  We will use the method of matched asymptotic expansions, solving separately in the near zone $r\ll R_L$ and the far zone $r\gg R_\star$, and matching the two in the overlap region $R_\star\ll r\ll R_L$.  More formally, the near (respectively, far) expansions are $\epsilon\to0$ fixing $R_\star$ (respectively, $R_L$), and we match the large-$r$ behavior of the near expansion to the small-$r$ behavior of the far expansion.

\subsection{Near-zone}

In the near-zone $r\ll R_L$, Eq.~\eqref{eq:StreamEquation} becomes
\begin{align}
	\label{eq:NearStreamEquation}
	r^2\pd_r\!\pa{\alpha^2\pd_r\psi}+\sin{\theta}\pd_\theta\!\pa{\frac{\pd_\theta\psi}{\sin{\theta}}}=0.
\end{align}
This is the vacuum magnetic stream equation in the Schwarzschild spacetime.  We will label its solutions with ``near.''  The general solution vanishing at both poles and falling off at large $r$ is given in a multipole expansion as (see App.~\ref{app:Harmonics})
\begin{align}
	\label{eq:NearPsi}
	\psi_{\rm near}(r,\theta)=\sum_{\ell=1}^\infty C_\ell R^>_\ell(r)\Theta_\ell(\theta).
\end{align}
The normalization of the radial functions is chosen so that $R^>_\ell(r)\to r^{-\ell}$ as $r\to\infty$.

To satisfy the boundary condition \eqref{eq:NearFluxBoundaryCondition} on the star, we must have
\begin{align}
	\psi_{\rm near}\big|_{r=R_\star}&=\psi_\star(\theta).
\end{align}
Provided the dipole moment $C_1=\mu$ is nonzero, the dipole term will always dominate at sufficiently large $r$, 
\begin{align}
	\label{eq:DipoleTerm}
	\psi_{\rm near}=\frac{\mu}{r}\sin^2{\theta},\quad r\to\infty.
\end{align}

\subsection{Far-zone}
\label{sec:FarZone}

\begin{figure}
\centering
\includegraphics[width=.4\textwidth]{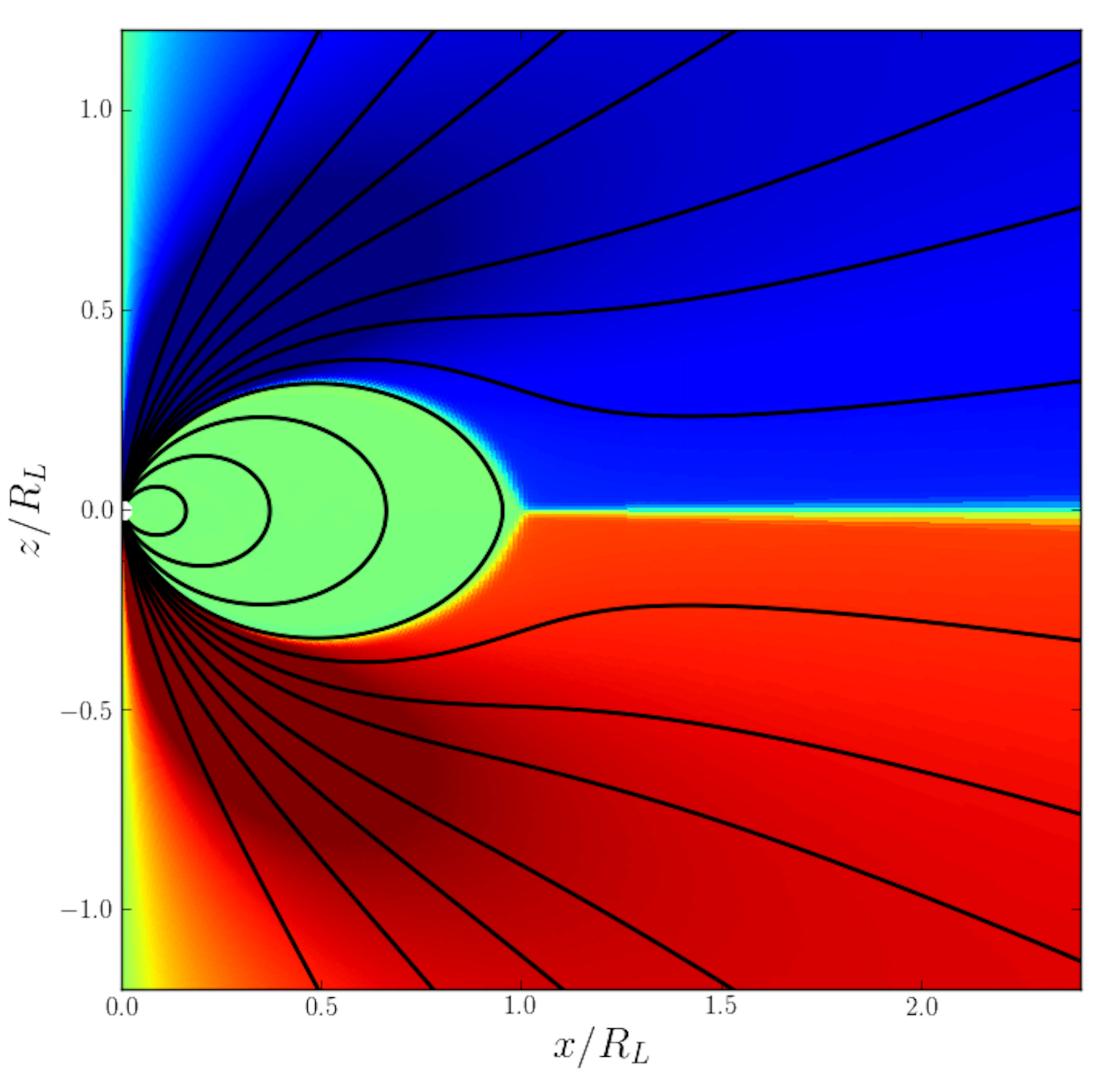}
\caption{Far-zone field configuration.  Poloidal field lines are shown as black and the color represents $I(\psi)$, with red to green to blue representing negative to zero to positive values.}
\label{fig:FarZoneDipole}
\end{figure}

\begin{figure}
\centering
\includegraphics[width=.45\textwidth]{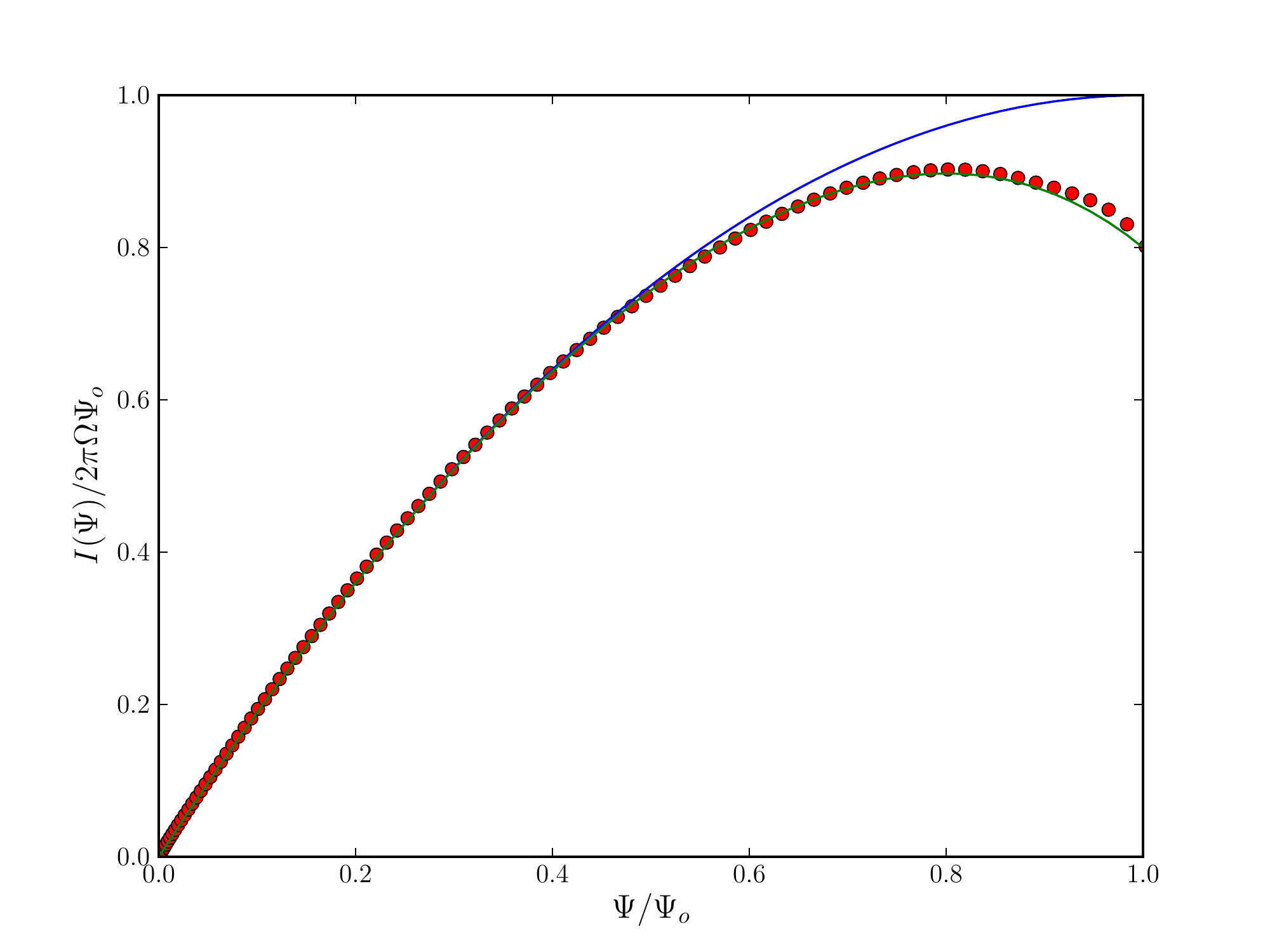}
\caption{Current distribution $I(\psi)$ (in the northern hemisphere). Red dots show the $\epsilon=1/50$ simulation result, the green line represents the fit \eqref{eq:DipoleSolution}, and the blue line shows the monopole distribution \eqref{eq:MichelCurrent}.  At each value of $\psi$, the fit differs from the numerical results by no more than $3\%$.}
\label{fig:Current}
\end{figure}

In the far-zone $r\gg R_\star$ (where also $r\gg M$), Eq.~\eqref{eq:StreamEquation} becomes
\begin{align}
	&r^2\pd_r\!\br{\pa{1-\Omega^2r^2\sin^2{\theta}}\pd_r\psi}
	\nonumber\\
	&\qquad+\sin{\theta}\pd_\theta\!\br{\frac{\pa{1-\Omega^2r^2\sin^2{\theta}}}{\sin{\theta}}\pd_\theta\psi}
	\label{eq:FarStreamEquation}\\
	&\qquad\qquad=-\pa{\frac{r}{2\pi}}^2I(\psi)I'(\psi).
	\nonumber
\end{align}
This is the stream equation in flat spacetime.  In general, it must be solved numerically.  We will label its solutions with ``far.''  To match the boundary condition \eqref{eq:FarFluxBoundaryCondition}, we must have asymptotically radial field lines
\begin{align}
	\label{eq:FarZoneLargeRadius}
	\psi_{\rm far}=A(\theta)+\mathcal{O}\!\pa{r^{-1}},
\end{align}
for some function $A(\theta)$ determined by solving the equations.  To match to the near solution \eqref{eq:DipoleTerm} in the overlap region, the solution must satisfy 
\begin{align}
	\label{eq:FarZoneSmallRadius}
	\psi_{\rm far}=\frac{\mu}{r}\sin^2{\theta},\quad r\to0.
\end{align}
The solution to Eq.~\eqref{eq:FarStreamEquation} satisfying Eqs.~\eqref{eq:FarZoneLargeRadius}-\eqref{eq:FarZoneSmallRadius} corresponds to the small-$\epsilon$ limit of the classic dipole pulsar problem in flat spacetime.  We use the force-free version of the HARM code \cite{Gammie03, Tchekhovskoy07} and set the inflow velocity into the current sheet to zero in order to minimize the numerical dissipation \cite{McKinney06}.  The simulation setup is described in Ref.~\cite{Tchekhovskoy16}.  Performing a sequence of simulations with decreasing $\epsilon$, we find good convergence by $\epsilon=1/50$.  The resulting flux function is shown in Fig.~\ref{fig:FarZoneDipole}.  The configuration is symmetric about the equatorial plane and the last open field line(s) have the value \cite{Timokhin06}
\begin{align}
	\label{eq:LastOpenFieldLine}
	\psi_o=n\mu\Omega,\qquad n\approx1.23.
\end{align}
There is no current flowing on the closed field lines,
\begin{align}
	I(\psi)=0,\quad\psi>\psi_o,
\end{align}
while the current on the open field lines is given to an excellent approximation (see Fig.~\ref{fig:Current}) by 
\begin{align}
	\label{eq:DipoleSolution}
	I(\psi)=\pm2\pi\Omega\psi\br{2-\frac{\psi}{\psi_o}-\frac{1}{5}\pa{\frac{\psi}{\psi_o}}^3},\quad\psi<\psi_o,
\end{align}
where we take the $+$ sign (respectively, $-$ sign) for the northern (respectively, southern) hemisphere.
 
The first two terms in the brackets are the current for the split monopole [Eq.~\eqref{eq:MichelCurrent}]; we see that the dipole provides a quartic correction.  This correction describes the volume return current, which flows close to the edge of the polar cap.  While Eq.~\eqref{eq:DipoleSolution} was obtained in the far expansion, it is independent of $r$ and hence holds everywhere.

\subsection{Stellar multipole moments}

To characterize the field on the star, it is useful to introduce stellar multipole moments $B_\ell$ by
\begin{align}
	\psi_\star(\theta)=R_\star^2\sum_{\ell=1}^\infty B_\ell\Theta_\ell(\theta).
\end{align}
The factor of $R_\star^2$ gives these moments units of magnetic field.  The stellar moments are related to the moments appearing in Eq.~\eqref{eq:NearPsi} by
\begin{align}
	B_\ell=\frac{C_\ell}{R_\star^{\ell+2}}\Delta_\ell(\C),
\end{align}
where $\Delta_\ell=R_\star^\ell\,R^>_\ell(R_\star)$ is a dimensionless function of $M$ and $R_\star$ (and hence depends only on the compactness $\C$).  The precise form of any $\Delta_\ell$ can be determined from the analysis of App.~\ref{app:Harmonics}.  For example, the dipole and quadrupole correction factors are
\begin{align}
	\Delta_1&=-\frac{3}{2\C^3}\br{\C\pa{\C+2}+2\log(1-\C)},\\
	\Delta_2&=-\frac{20}{3\C^2}\br{4-\pa{4-3\C}\Delta_1}.
\end{align}
By construction, all the $\Delta_\ell$ satisfy
\begin{align}
	\lim_{\C\to0}\Delta_\ell=1.
\end{align}
At $\Delta_\ell=1$ the source and field moments are related as they are in flat spacetime, so we may view $\Delta_\ell$ as the fractional correction due to general relativity.

\subsection{Range of validity}

We have assumed that the dipole moment $C_1=\mu$ is nonzero and included only that term in the overlap region [see Eqs.~\eqref{eq:DipoleTerm}-\eqref{eq:FarZoneSmallRadius}].  This means that our approximation will only be reliable when the field is predominantly dipole in the overlap region $R_{\star}\ll r\ll R_L$.  To be safe, the approximation is valid if $C_1\gg C_\ell/R_L^{\ell-1}$ for $\ell\geq2$.  Since $\Delta_\ell$ is order unity, $B_\ell\sim C_\ell/R_\star^\ell$.  Thus, the complete conditions for the validity of our approximation are
\begin{align}
	\epsilon\ll1,\qquad\frac{B_\ell}{B_1}\ll\frac{1}{\epsilon^{\ell-1}},\quad\ell\geq2.
\end{align}
The latter condition is violated only when higher stellar moments dominate the dipole by rather extreme amounts.  If this is the case, then the general method still applies, but one should instead assume the dominant moment in Eqs.~\eqref{eq:DipoleTerm} and \eqref{eq:FarZoneSmallRadius}.

\subsection{Power output}

The electromagnetic luminosity (energy loss rate) of the star is \cite{Beskin10}
\begin{align}
	\mathcal{L}=2\Omega\int_0^{\psi_o}I(\psi)\ed\psi.
\end{align}
Plugging in Eqs.~\eqref{eq:DipoleSolution} and \eqref{eq:LastOpenFieldLine} yields
\begin{align}
	\label{eq:Power}
	\mathcal{L}=4\pi\frac{47}{75}n^2\mu^2\Omega^4=0.95\times\pa{\sqrt{4\pi}\mu}^2\Omega^4.
\end{align}
Note that the magnetic moment is $\hat{\mu}=\sqrt{4\pi}\mu$ in Gaussian units.  Previous studies for $\epsilon\sim0.2$ in flat spacetime have found the power to be $\approx1\times\hat{\mu}^2\Omega^4$ \cite{Contopoulos99, Gruzinov05, McKinney06, Timokhin06}, which is consistent with the small-$\epsilon$ result up to numerical accuracy.

In curved spacetime, $\mu$ is the magnetic moment inferred from the field in the overlap region (far from both the star and the light cylinder).  We therefore prove that, when compared at fixed $\mu$, there are \textit{no} GR corrections at leading order in $\epsilon$.  The corrections found at fixed $\mu$ in Refs.~\cite{Ruiz14,Petri16} will therefore disappear at smaller values of $\epsilon$.

In terms of the stellar dipole moment $B_1$, Eq.~\eqref{eq:Power} becomes
\begin{align}\label{eq:L}
	\mathcal{L}=0.95\times\pa{\frac{\sqrt{4\pi}B_1R_\star^3}{\Delta_1(\C)}}^2\Omega^4.
\end{align}
This formula is rather insensitive to the details of the stellar magnetic field, depending only on its dipole component $B_1$.  From an observation of pulsar spindown, we can therefore infer the dipole field $B_1$.  Noting that the angular momentum $J$ of the star is given by $J=MR_\star^2\I\Omega$ and using the general relation $dE/dt=\Omega dJ/dt$, to leading order in $\epsilon$ we have
\begin{align}
	\label{eq:B1}
	\sqrt{4\pi}B_1=\frac{\Delta_1(\C)}{2\pi}\sqrt{\frac{1}{0.95}\frac{\I M}{R_\star^4}P\dot{P}},
\end{align}
where $P=2\pi/\Omega$ and $\dot{P}>0$ are the period and period derivative, respectively.  We can view the factor of $\Delta_1(\C)$ as the fractional correction due to general relativity.  At a typical compactness, we have $\Delta_1(1/2)=1.64$, so GR makes a roughly $60\%$ correction. 

These results pertain to force-free fields, showing that electromagnetic losses decrease with increasing compactness (at fixed surface magnetic field strength).  Interestingly, the opposite is true for vacuum fields (for the oblique rotator): the losses \textit{increase} with increasing compactness \cite{Rezzolla04}.  The same reversal occurs for losses from stellar pulsations \cite{Abdikamalov09}.

\subsection{Polar Caps}
\label{sec:PolarCaps}

In the classic dipole pulsar, current flows off the star only in regions near the poles called the ``polar caps.''  When higher moments are allowed, there can be multiple regions of current outflow and they can be located anywhere on the star.  We will show, however, that to leading order in $\epsilon$, there are always exactly two regions, connecting to the northern and southern outflows in the asymptotic region.  We will continue to refer to these regions as northern and southern polar caps, although they need not be located at the poles.

The regions of current outflow (open field lines) are described by $0<\psi<\psi_o$ in the far-zone.  Respectively, the bounding field lines $\psi=0$ and $\psi=\psi_o$ intersect the star at angles $\theta_0$ and $\theta_c$ determined by [see Eq.~\eqref{eq:LastOpenFieldLine}]
\begin{subequations}
\label{eq:PolarCaps}
\begin{align}
	\psi_\star(\theta_0)&=0,\\
	\psi_\star(\theta_c)&=B_1R_\star^2\frac{n}{\Delta_1}\epsilon.
\end{align}
\end{subequations}
Since Eqs.~\eqref{eq:PolarCaps} are to be solved to leading order in $\epsilon$, all solutions for $\theta_c$ are parametrically near a solution for $\theta_0$.  The leading Taylor approximation for $\psi_\star(\theta_c)$ about the nearby $\theta_0$ is 
\begin{align}
	\psi_\star(\theta_c)\approx\frac{\pa{\theta_c-\theta_0}^m}{m!}\psi^{(m)}(\theta_0),
\end{align}
where $\psi^{(m)}$ is the first nonvanishing derivative of $\psi$ at $\theta_0$.  Thus, the leading-order solution for $\theta_c$ is
\begin{align}
	\label{eq:CriticalAngles}
	\theta_c=\theta_0+\pa{m!\frac{B_1R_\star^2}{\psi_\star^{(m)}(\theta_0)}\frac{n}{\Delta_1}\epsilon}^{1/m}.
\end{align}
If $m$ is even, then there are two solutions, corresponding to the $\pm$ ambiguity in the power of $1/m$.  This associates up to two values of $\theta_c$ with each root $\theta_0$.

Among this potential panoply of solutions to Eqs.~\eqref{eq:PolarCaps}, only the values of $\theta_c$ closest to the poles actually connect to the far-zone outflow.\footnote{In the overlap region viewed from the near expansion, the field is dipolar but $\psi_o$ is parametrically small and hence, the last open field line is actually parametrically close to the pole.  This means that the near-zone outflow is parametrically close to a field line that is asymptotically polar as $r\to0$ in the near-zone, which corresponds to the solution to Eqs.~\eqref{eq:PolarCaps} for which $\theta_c$ is closest to the pole.  For the northern flow, $\theta_c>\theta_0$, while for the southern flow, $\theta_c<\theta_0$.}  In particular, the northern polar cap corresponds to $\theta_0<\theta<\theta_c$ for the smallest allowed $\theta_c$ and its associated $\theta_0$, while the southern polar cap corresponds to $\theta_c<\theta<\theta_0$ for the largest allowed $\theta_c$ and its associated $\theta_0$.  To summarize, the northern/southern polar cap is found by finding all zeros $\theta_0$ of $\psi_\star(\theta)$, computing associated $\theta_c$ values from Eq.~\eqref{eq:CriticalAngles}, and choosing the pairs $\{\theta_0,\theta_c\}$ containing the smallest/largest values of $\theta_c$.

Notice that the width $\ab{\theta_c-\theta_0}$ of the polar cap region is set by the order $m$ of the first nonvanishing derivative at $\theta_0$.  When a polar cap lies at a pole $\theta=0$ or $\theta=\pi$, the order is always at least $m=2$, since all physical flux functions vanish quadratically on the axis.  Generically, it will be exactly $m=2$.  On the other hand, if the polar cap is shifted, there is no reason to expect the first derivative to vanish, and generically we will have $m=1$.  Thus, apart from finely tuned situations,
\begin{align}
	\textrm{polar cap width}\sim
	\begin{cases}
		\sqrt{\epsilon}&\textrm{if on rotational axis},\\
		\epsilon&\textrm{if displaced}.
	\end{cases}
\end{align}
Note that while displaced polar caps are narrower, the total surface area is comparable since $\pa{\theta_c-\theta_0}\sin\theta_c\sim\epsilon$ in both cases.

We have defined the polar cap as the angular region of current flow on the stellar surface.  More generally, we could consider the angular region of the current flow at an arbitrary radius.  The analysis of the section still applies using the flux function at that radius.  As the radius increases and the dipole term starts to dominate, the polar cap will eventually move to the pole and assume the corresponding $\sqrt{\epsilon}$ scaling, as indeed it must in order to match to the far-zone solution (Fig.~\ref{fig:FarZoneDipole}).  The transition region between the scalings can be handled by the methods discussed in App.~\ref{app:Quadrudipole} for the quadrudipole pulsar.

\subsection{Near-zone Current}

We can compute the charge-current $J^\mu=\nabla_\nu F^{\mu\nu}$ from Eqs.~\eqref{eq:StarMetric}, \eqref{eq:F} and \eqref{eq:OmegaField}.  To leading order $\mathcal{O}(\epsilon)$ in the near zone $r\ll R_L$ (i.e., fixing $R_\star$), we find
\begin{subequations}
\label{eq:Current}
\begin{align}
	J^t&=\frac{2\pa{\Omega-\Omega_Z}}{r\pa{r-2M}}\br{\pa{r-3M}\pd_r\psi+\cot{\theta}\pd_\theta\psi},\\
	J^r&=-\frac{\pd_\theta I(\psi)}{2\pi r^2\sin{\theta}}=-\frac{\pd_\theta I(\psi)}{2\pi\sqrt{-g}},\\
	J^\theta&=\frac{\pd_r I(\psi)}{2\pi r^2\sin{\theta}}=\frac{\pd_r I(\psi)}{2\pi\sqrt{-g}},\\
	J^\phi&=0,
\end{align}
\end{subequations}
where a near-zone solution \eqref{eq:NearPsi} should be used for $\psi$.  Note that the current $I$ is proportional to $\Omega$ by the boundary condition \eqref{eq:FarCurrentBoundaryCondition} and in any case by the explicit formula \eqref{eq:DipoleSolution}.  Eqs.~\eqref{eq:Current} are then seen to be $\mathcal{O}(\epsilon)$, noting that $\Omega=\epsilon R_\star$ and $R_\star$ is fixed in the near expansion.  GR effects for the charge density $J^t$ were also discussed in Refs.~\cite{Beskin90, Muslimov92}.

The norm of the current is nonzero only at quadratic order $\mathcal{O}\!\pa{\epsilon^2}$, where we obtain
\begin{align}
	\label{eq:CurrentNorm}
	J^2&=-\frac{4\pa{\Omega-\Omega_Z}^2}{r^3\pa{r-2M}}\br{\pa{r-3M}\pd_r\psi+\cot{\theta}\pd_\theta\psi}^2\\
	&\qquad+\frac{r}{r-2M}\br{\frac{\pd_\theta I(\psi)}{2\pi r^2\sin{\theta}}}^2+r^2\br{\frac{\pd_r I(\psi)}{2\pi r^2\sin{\theta}}}^2.
	\nonumber
\end{align}
The current is only nonzero in two parametrically small regions whose footprints on the star are called the northern and southern polar caps (see Sec.~\ref{sec:PolarCaps}).  Thus, in each region there is a small parameter to make further simplifications to the expressions \eqref{eq:Current}-\eqref{eq:CurrentNorm} for the current.   

\section{Dipole Pulsar}
\label{sec:Dipole}

\begin{figure}
\centering
\includegraphics[width=.4\textwidth]{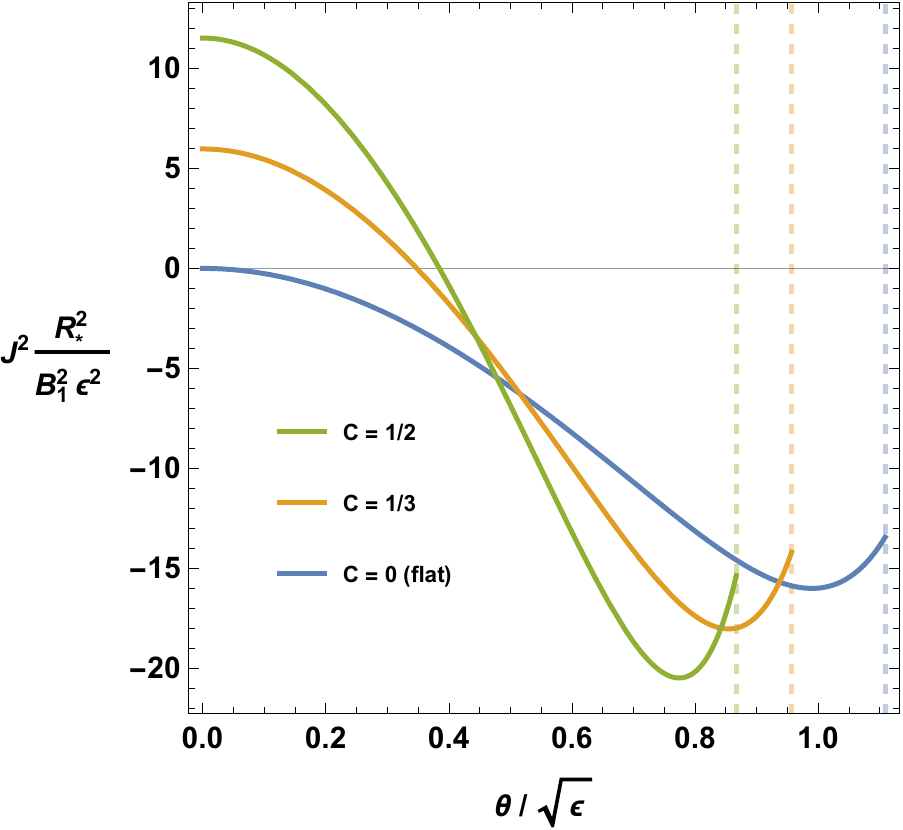}
\caption{Norm of the current on the polar cap of the dipole pulsar.  The vertical lines delimit the edge of cap, where the current shuts off.  We choose $\I=2/5$ and show three different values of stellar compactness $\C$.  We see that spacelike current is generic at any reasonable values of compactness.  The case $\C=0$ corresponds to ignoring the effects of GR, in which case the spacelike current disappears.}
\label{fig:DipoleCurrentNorm}
\end{figure}

Suppose that the field on the star is pure dipole,
\begin{align}
	\label{eq:DipoleFlux}
	\psi_\star(\theta)=B_1R_\star^2\sin^2{\theta}.
\end{align}
Then the near-zone solution is
\begin{align}
	\label{eq:NearDipole}
	\psi_{\rm near}(r,\theta)=\mu R^>_1(r)\sin^2{\theta},
\end{align}
where $\mu=C_1=B_1 R_\star^3/\Delta_1$ and $R^>_1(r)$ is given in Eq.~\eqref{eq:R1}.  The polar caps are found by solving Eqs.~\eqref{eq:PolarCaps} using Eq.~\eqref{eq:DipoleFlux},
\begin{align}
	\sin^2{\theta_0}=0,\qquad\sin^2{\theta_c}=\frac{n}{\Delta_1}\epsilon,
\end{align}
which gives, to leading order,
\begin{align}
	\theta_N\in\pa{0,\theta_*},\qquad\theta_S\in\pa{\pi-\theta_*,\pi},
\end{align}
where we defined
\begin{align}
	\theta_*=\sqrt{\frac{n}{\Delta_1}\epsilon}.
\end{align}
The entire magnetosphere is reflection-symmetric, so we may focus on the northern cap for simplicity.  The current norm $J^2$ on the surface may be obtained from Eq.~\eqref{eq:CurrentNorm} by plugging in Eq.~\eqref{eq:DipoleSolution} and Eq.~\eqref{eq:NearDipole}.  On the north cap, noting that $\theta<\theta_o\sim\sqrt{\epsilon}$ and keeping to leading order in $\epsilon$, the result is
\begin{align}
	\label{eq:NorthDipoleCurrentNorm}
	J^2\big|_{r=R_\star}=\pa{\frac{B_1}{R_\star}}^2f\!\pa{\I,\C,\frac{\theta}{\theta_*}}\epsilon^2,
\end{align} 
with 
\begin{align}
	f(\I,\C,y)=\frac{16}{1-\C}\br{\pa{\tfrac{2}{5}y^6+y^2-1}^2-\pa{1-\I\C}^2}.
\end{align}
The current norm is plotted for some representative values of $\C$ and $\I$ in Fig.~\ref{fig:DipoleCurrentNorm}.  On the pole $y=0$, we have
\begin{align}
	\label{eq:NorthPoleDipoleCurrentNorm}
	f\big|_{\rm pole}=16\I\C\pa{\frac{2-\I\C}{1-\C}}
	=\frac{16}{\alpha^2}\pa{2-\frac{\Omega_Z}{\Omega}}\frac{\Omega_Z}{\Omega},
\end{align}
where for the second equality we have used Eq.~\eqref{eq:AlphaOmega}.  Since $\C<1$ and $\I \C<2$ for any reasonable star, we see that $f$ is always positive on the pole.  This means that the current is always spacelike in a region near the pole.  In light of the appearance of $\Omega_Z$ in Eq.~\eqref{eq:NorthPoleDipoleCurrentNorm}, we can ascribe the spacelike current to the effects of frame-dragging \cite{Philippov15b}.

Ignoring the effects of gravity corresponds to taking $\C=0$.  In this case, $f=0$ on the pole and it is straightforward to prove that $f<0$ everywhere else, as seen in the plot.  This explains why work in flat spacetime missed the presence of spacelike current.

Concurrently with our work, Ref.~\cite{Belyaev16} gave an independent analytical treatment of the current norm $J^2$ in the polar cap of a slowly rotating dipole pulsar in general relativity.  Their method uses the monopole current \eqref{eq:MichelCurrent} along with the dipole flux function \eqref{eq:NearDipole}.  By physical reasoning and comparison with numerical simulations, the authors recognized that this approach should only be reliable near the pole.  Our result \eqref{eq:DipoleSolution} for the dipole current quantifies the associated error, showing that their $J^2$ is valid to second order in $\theta/\theta_c$.  The error for $\theta\sim\theta_c$ is visible in their Fig.~1 (analogous to our Fig.~\ref{fig:DipoleCurrentNorm}), which lacks the upturn seen in Fig.~\ref{fig:DipoleCurrentNorm}.  Our work also goes beyond that of Ref.~\cite{Belyaev16} by systematizing the approximation method in terms of matched asymptotic expansions, considering higher multipole moments, and addressing the spindown luminosity.

We may also compare our results for the stellar current flow with previous force-free numerical simulations \cite{Philippov15b,Belyaev16}.  These simulations were performed at relatively larger values of $\epsilon\sim1/5$, so there is no a priori reason to expect agreement with our perturbative calculation.  Nevertheless, there is full qualitative agreement of all features: spacelike current near the pole, transitioning to a timelike outflow, and finally, a timelike volume return current near the edge of the cap.  This suggests that $\epsilon\sim1/5$ numerical simulations can be used to model the regime $\epsilon\ll1$ relevant to most pulsars, and conversely that our results may still be applicable for fast rotation.

\section{Quadrudipole Pulsar}
\label{sec:Quadrudipole}

\begin{figure*}
\centering
\subfigure[\ Stellar flux function]{\label{fig:StellarFlux}
\includegraphics[width=.29\textwidth]{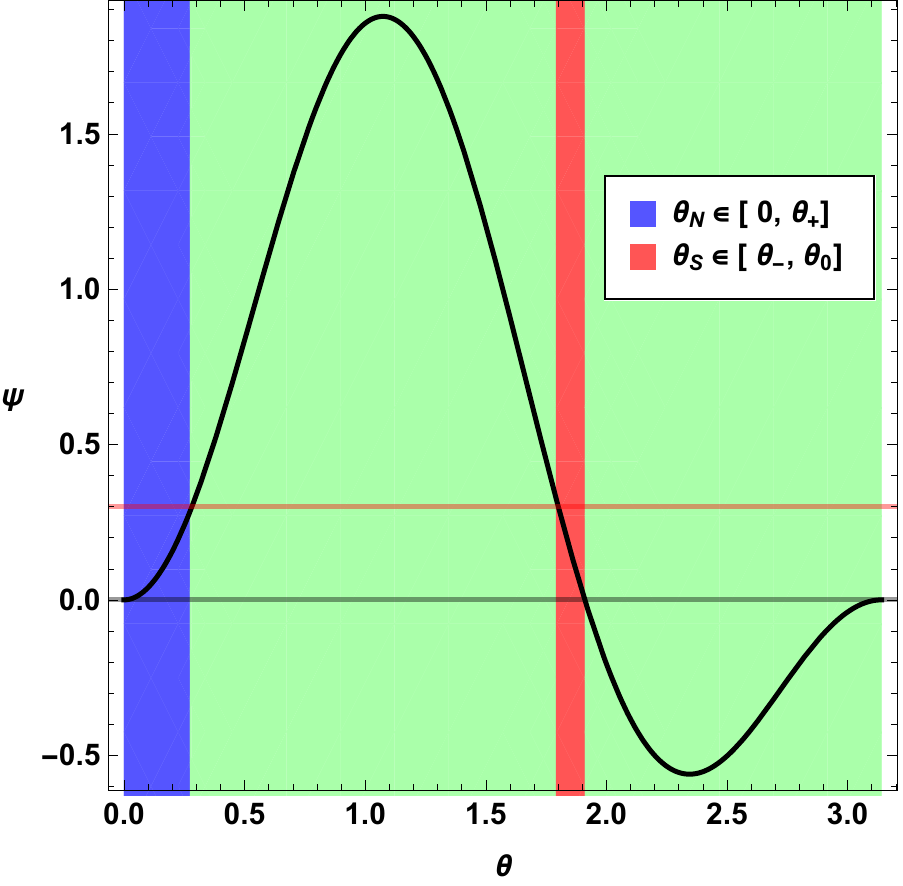}}
\quad
\subfigure[\ Northern cap current]{\label{fig:QuadrupoleNorthCap}
\includegraphics[width=.31\textwidth]{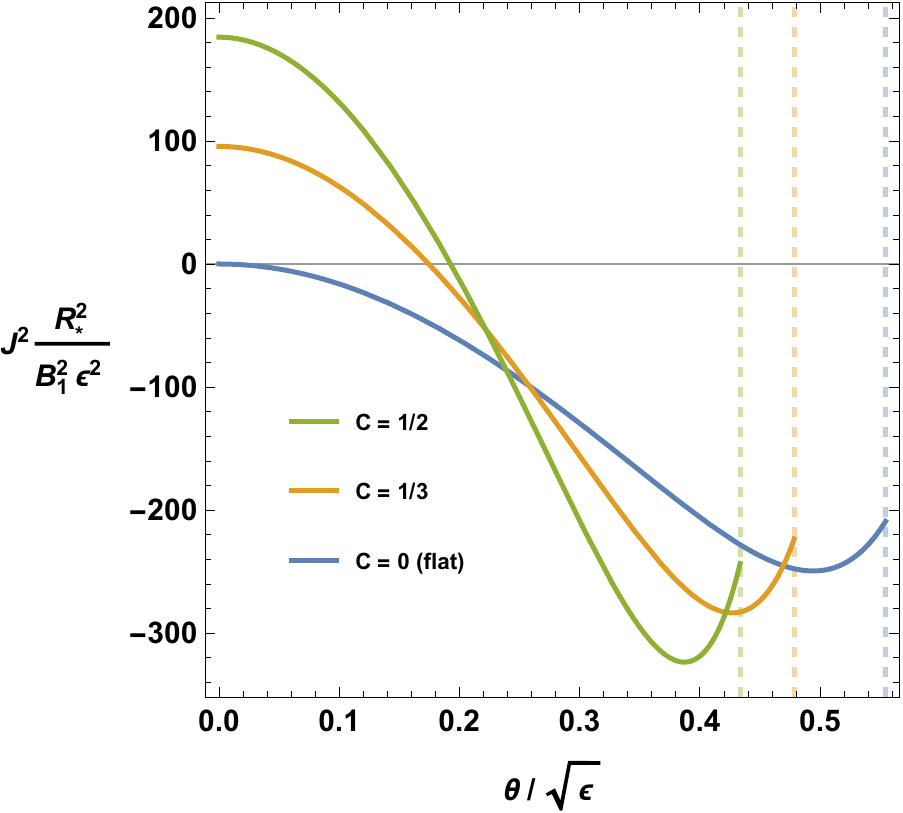}}
\quad
\subfigure[\ Southern cap current]{\label{fig:QuadrupoleSouthCap}
\includegraphics[width=.31\textwidth]{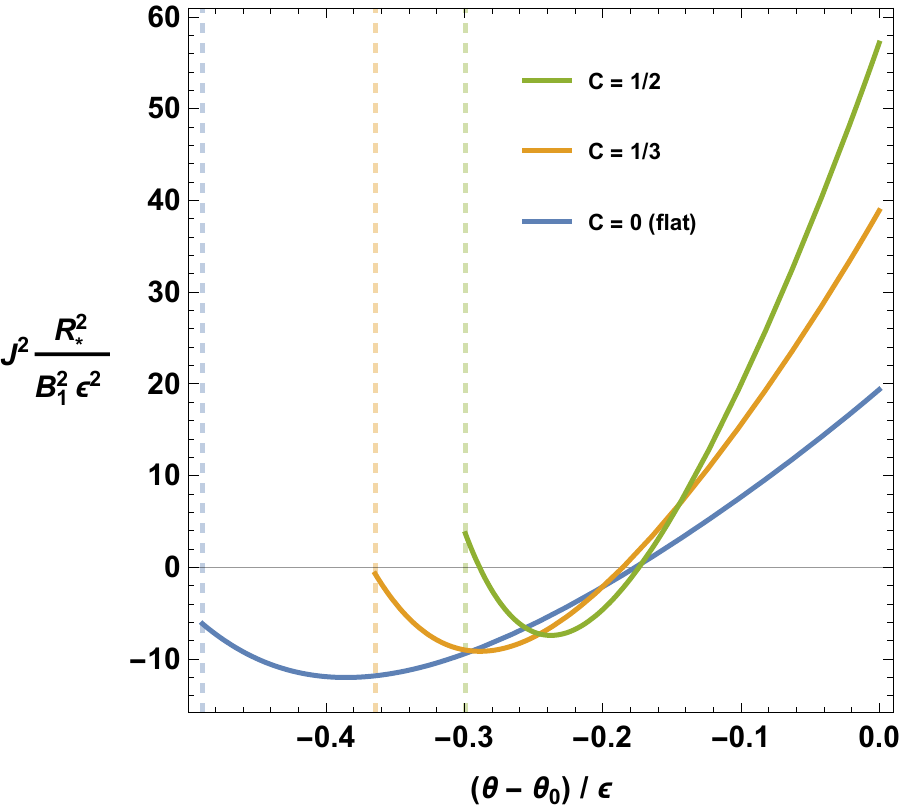}}\\
\subfigure[\ Near-zone zoomed in]{\label{fig:NearNearZone}
\includegraphics[width=.3\textwidth]{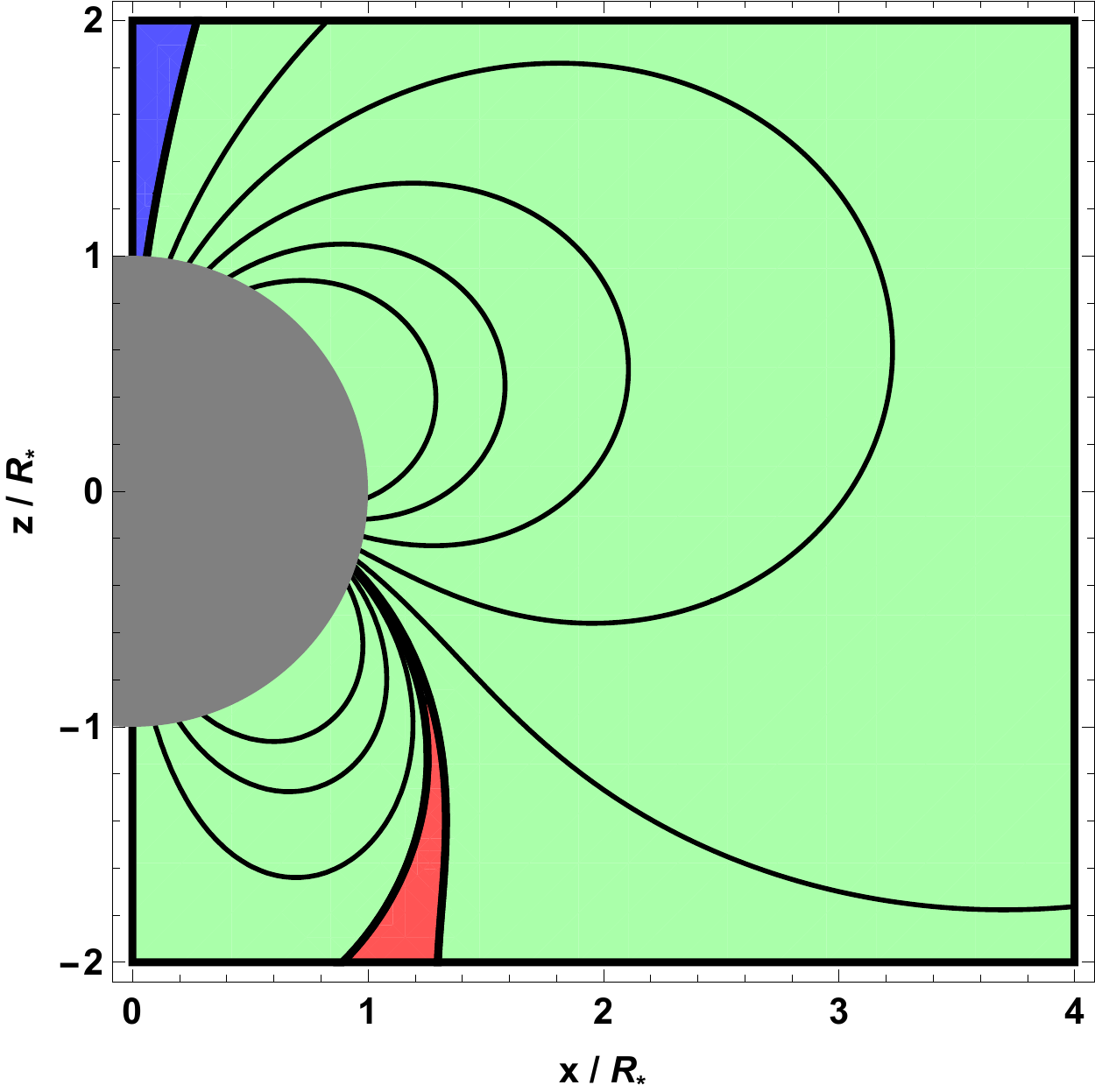}}
\quad
\subfigure[\ Near-zone zoomed out]{\label{fig:FarNearZone}
\includegraphics[width=.3\textwidth]{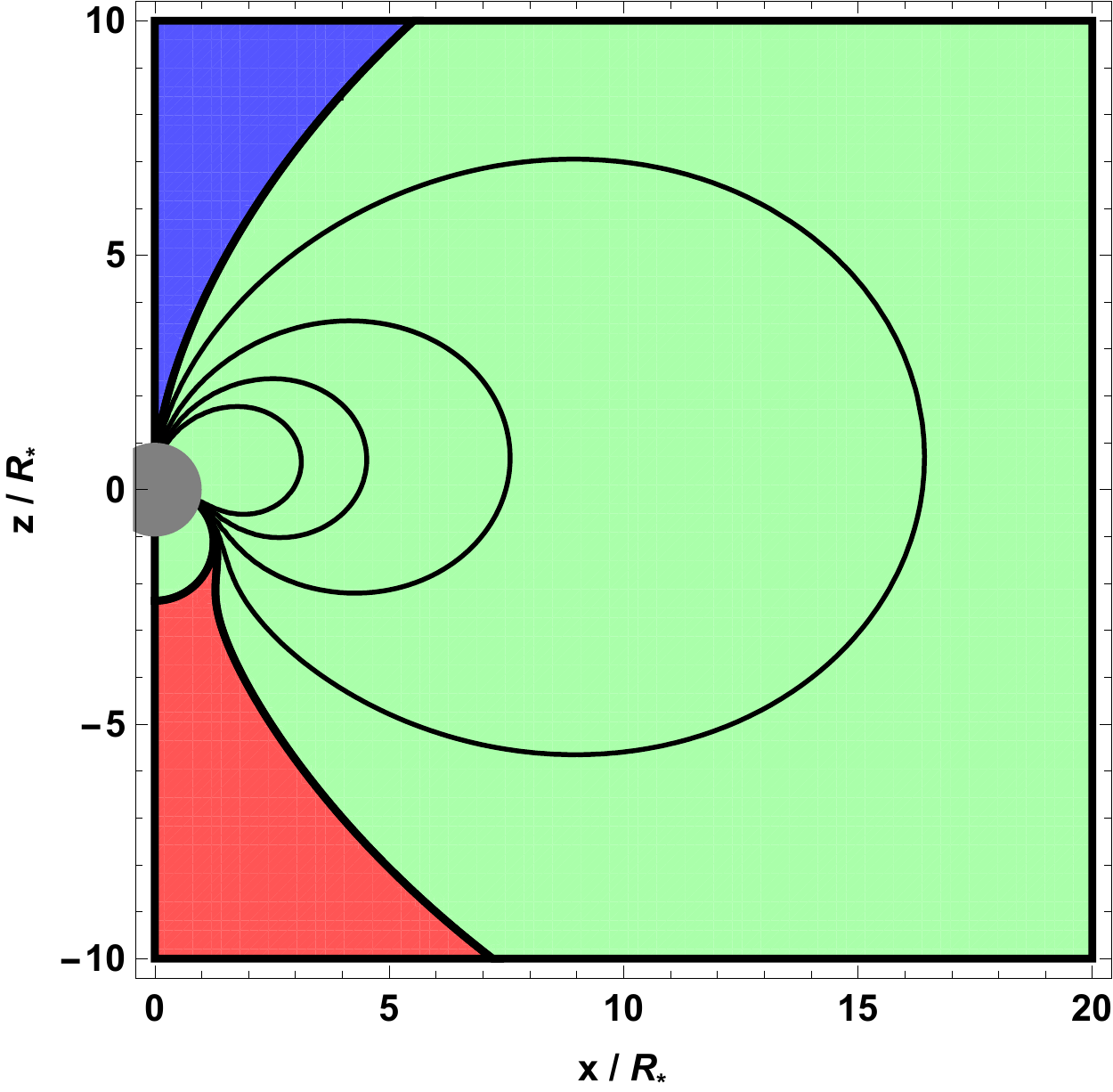}}
\quad
\subfigure[\ Far-zone ]{\label{fig:FarZoneDipoleBis}
\includegraphics[width=.3\textwidth]{FarZoneDipole}}
\caption{Quadrudipole pulsar.  We consider a stellar magnetic field with quadrupole and dipole moments in the ratio 3:1 (i.e., $q=3$).  The stellar flux function is shown in \subref{fig:StellarFlux}.  The horizontal line is the flux value of the last open field line(s), which is proportional to $\epsilon$ (here taken to be large for clarity) and defines the northern and southern polar caps (blue and red regions).  The associated region of current outflow is correspondingly colored in the near-zone field structure \subref{fig:NearNearZone} and \subref{fig:FarNearZone}, where we use $\epsilon=1/50$.  In \subref{fig:FarZoneDipoleBis}, we show the far-zone simulation result (identical to Fig.~\ref{fig:DipoleCurrentNorm}) as well.  Finally, \subref{fig:QuadrupoleNorthCap} and \subref{fig:QuadrupoleSouthCap} show the norm of the current in the polar caps, which is spacelike over large regions for reasonable values of compactness.  The annular shape and parametrically small width ($\sim\epsilon$ instead of $\sqrt{\epsilon}$) of the southern polar cap, together with the high magnetic field curvature, make it a favorable region for pair production and may help explain the modified beam characteristics of millisecond pulsars.}
\label{fig:Quadrudipole}
\end{figure*}

We now consider a pulsar with both a dipole and a quadrupole moment,
\begin{align}
	\label{eq:QuadrudipoleFlux}
	\psi_\star(\theta)&=R_\star^2\pa{B_1\sin^2{\theta}+B_2\cos{\theta}\sin^2{\theta}}\\
	&=B_1R_\star^2\pa{1+q\cos{\theta}}\sin^2{\theta},
\end{align}
where we defined the ratio $q$ between the moments,
\begin{align}
	q=\frac{B_2}{B_1}.
\end{align}
This configuration was studied previously in Ref.~\cite{Barnard82}, without constructing a self-consistent global model.  Our approximation is valid when $\epsilon\ll1$ and $q\ll\epsilon^{-1}$.  The near-zone flux function is
\begin{align}
	\psi_{\rm near}=C_1R^>_1(r)\sin^2{\theta}+C_2R^>_2(r)\cos{\theta}\sin^2{\theta},
\end{align}
where $C_1=B_1R_\star^3/\Delta_1$ and $C_2=B_2R_\star^4/\Delta_2$.  Expressions for $R_1^>(r)$ and $R_2^>(r)$ are given in Eqs.~\eqref{eq:R1} and \eqref{eq:R2}.  

Since $q\to-q$ is equivalent to reflection about the equatorial plane, we may assume $q>0$ without loss of generality.  The polar caps are found from Eqs.~\eqref{eq:PolarCaps} using Eq.~\eqref{eq:DipoleFlux},
\begin{subequations}
\label{eq:QuadrudipolePolarCaps}
\begin{align}
	\sin^2{\theta_0}\pa{1+q\cos{\theta_0}}&=0,\\
	\sin^2{\theta_c}\pa{1+q\cos{\theta_c}}&=\frac{n}{\Delta_1}\epsilon,
\end{align}
\end{subequations}
according to the method described in Sec.~\ref{sec:PolarCaps}.

For $q<1$, the situation is similar to the dipolar case.  The only solutions for $\theta_0$ are the poles $0$ and $\pi$, and the associated polar caps are
\begin{align}
	q<1:\qquad\theta_N\in\pa{0,\theta_+},\quad\theta_S\in\pa{\theta_-,\pi},
\end{align}
where 
\begin{align}
	\label{eq:SmallQuadrudipole}
	\theta_+=\sqrt{\frac{n\epsilon}{\pa{1+q}\Delta_1}},\qquad\theta_-=\pi-\sqrt{\frac{n\epsilon}{\pa{1-q}\Delta_1}}.
\end{align}
On the other hand, for $q>1$ there is a new zero of the flux function located at
\begin{align}
	\theta_0=\pi-\arctan{\sqrt{q^2-1}},
\end{align}
and the polar caps are
\begin{align}
	q>1:\qquad\theta_N\in\pa{0,\theta_+},\quad\theta_S\in\pa{\theta_-,\theta_0},
\end{align}
where now
\begin{align}
	\label{eq:LargeQuadrudipole}
	\theta_-=\theta_0-\frac{q^2}{\pa{q^2-1}^{3/2}}\frac{n}{\Delta_1}\epsilon.
\end{align}
Thus, the northern polar cap is similar, but the southern cap is a narrow annulus ($\sim\epsilon$ instead of $\sqrt{\epsilon}$) located away from the pole.  The $q<1$ and $q>1$ results both break down for $\ab{1-q}\lesssim\epsilon$.  This regime seems too highly tuned to be important in nature, but for completeness we discuss it in App.~\ref{app:Quadrudipole}.  All three regimes can be handled simultaneously by working with exact solutions of Eqs.~\eqref{eq:QuadrudipolePolarCaps}, which we also provide in App.~\ref{app:Quadrudipole}.

The current norm $J^2$ may be obtained from Eqs.~\eqref{eq:CurrentNorm} and \eqref{eq:QuadrudipoleFlux}.  We give the full expression in Eq.~\eqref{eq:QuadrudipoleCurrentNorm}.  We can expand in $\epsilon$ in each cap of each regime of $q$ separately to produce analogs of Eq.~\eqref{eq:NorthDipoleCurrentNorm}, but the expressions are not transparent.  Instead, we evaluate on the edge of the polar caps, where $\psi=0$, and consider various limits.

First, for $q>1$ in the flat limit $\mathcal{C}\to0$, we have
\begin{align}
	J^2_N\big|_{\rm flat}&=0,\\
	J^2_S\big|_{\rm flat}&=\pa{\frac{2B_1}{R_\star}}^2\pa{q^2+\frac{3}{\Delta_1}}^2\pa{1-\frac{1}{q^2}}\frac{\epsilon^2}{q^2}.
\end{align}
Here, $J^2_N$ is the value at $r=R_\star$ and $\theta=0$, while $J^2_S$ is evaluated at $R=R_\star$ and $\theta=\theta_0$.  The northern current is null on the pole, while the southern current is always positive.  This shows that the current can be spacelike even when gravity is ignored.  Thus, while frame-dragging does produce spacelike current, in general it is not the only effect involved.  The full expression for $J^2$ on the southern cap $\theta_0$ is given in Eq.~\eqref{eq:QuadrudipoleCurrentNormPolarCap}.

When $q<1$, the edges of the caps are the poles and
\begin{align}
	J^2_{N/S}=\pa{\frac{B_1}{R_\star}}^2\br{16\I\C\pa{\frac{2-\I\C}{1-\C}}\pa{1\pm q^2}}\epsilon^2.
\end{align}
Here, $J^2_N$ is the value at $r=R_\star$ and $\theta=0$ and corresponds to the plus sign, while $J^2_S$ is evaluated at $R=R_\star$ and $\theta=\pi$ and corresponds to the minus sign.  In particular, we see that the dipole formula \eqref{eq:NorthPoleDipoleCurrentNorm} is corrected by a factor of of $1\pm q^2$.

While we have focused on the fields at the stellar surface, the analysis of the angular extent of the current flow regions generalizes to arbitrary radius by replacing $q=B_2/B_1$ with the corresponding moment ratio computed at that radius.  This defines a function $q(r)$ that decreases roughly as $1/r$.  Near the star where $q>1$, the width of the southern outflow scales as $\epsilon$, while further away it scales as $\sqrt{\epsilon}$.  The scaling in the transition regime $q\sim1$ is discussed in App.~\ref{app:Quadrudipole}, which also gives a master expression valid for all $q$.

Fig.~\ref{fig:Quadrudipole} illustrates our results for the quadrudipole pulsar with $q>1$.  We have also confirmed the overall field topology by running a numerical simulation in flat spacetime.

\section{More General Models}
\label{sec:Models}

We now discuss more general magnetic field configurations, including two that have arisen in recent research on neutron star magnetic field structure.  We first consider the superposition of a dipole and octupole, 
\begin{align}
	\psi_\star(\theta)=B_1R_\star^2\br{1+\hat{q}\pa{1-5\cos^2{\theta}}}\sin^2{\theta}.
\end{align}
It is straightforward to apply the method, but the formulas are complicated and we do not give details.  We discuss the case $\hat{q}=-1/3$, which was was found to be an attractor state in Hall-MHD studies of the magnetic field evolution in the neutron star crust \cite{Gourgouliatos14}.  In this case, the polar caps are both at the poles, where there are regions of spacelike current.  That is, the Hall attractor pulsar is qualitatively similar to the dipole pulsar.

On the other hand, if $\hat{q}>1$, then the polar caps are shifted away from the poles.  By including a quadrupole moment as well, we could arrange to place the caps anywhere.  Including even higher moments introduces a more complicated field structure in the closed zones, but does not change the overall topology.  This is the limit of qualitatively new features accessible at small $\epsilon$, so while it would be tempting to dramatize the power of the method by including higher moments, we see no reason beyond amusement to pursue the hexadecaquadrudipoctupole. 

Another recent study suggests that the first several moments should be all roughly the same order of magnitude \cite{Geppert14}.  In this case, the position of the polar caps is essentially random.  However, the power output is still fixed by the dipole moment $B_1$.

\section{Discussion}
\label{sec:Discussion}

In this paper, we developed an analytical technique to study the aligned force-free magnetosphere in the limit of slow stellar rotation with a generic magnetic field on the stellar surface, including all effects of general relativity.  Previous numerical models were limited to pulsars with fast rotation (e.g., $\epsilon\sim0.2$) and dipolar surface fields.  Our analytical results cover the regime $\epsilon\to0$ of interest for most pulsars and extend to more general surface fields.  Our only restriction on the magnetic field is that the dipole moment dominates far from the star (but still well inside the light cylinder). This assumption holds unless higher moments dominate the dipole by extreme amounts on the stellar surface.

We showed that the spindown luminosity is completely determined by the dipolar component of the surface magnetic field.  This means that pulsar magnetic field strengths typically quoted from spindown measurements correspond only to the dipolar component.  If pulsar magnetic fields contain many moments in equipartition \cite{Geppert14}, then the quoted value could differ from the true surface field strength by an order of magnitude or more.  We also showed that ignoring GR effects leads to a roughly 60\% error in the estimated surface field strength.  This occurs because the radial dependence of the dipolar magnetic field near the star is modified from the perfect $1/r^3$ scaling of flat spacetime.  The corrected formula for the dipolar component inferred from spindown is given in Eq.~\eqref{eq:B1}, where $B_1$ is one-half the value on the magnetic pole of a pure dipole.  This formula will receive an additional order-unity correction when inclination is taken into account.

Our results show that aligned GR force-free magnetospheres generically have a region of spacelike current in the polar caps, which favors the existence of $e^-/e^+$ discharges \cite{Timokhin13}.  This is generally consistent with the first-principles kinetic simulation of Ref.~\cite{Philippov15b}.  One difference is that these authors also found a lower limit for the value of stellar compactness at which pair production can occur, whereas the force-free solution has spacelike current at any compactness.  The discrepancy was attributed to lower values of the bulk current in the polar cap, a feature ultimately traced to the existence of a large vacuum region near the equatorial current sheet, where the volume return current flows in the force-free solution.  If this gap is filled with plasma, then the polar cap current increases and reaches its force-free value \cite{Philippov15b}.  The question of populating this region with pair plasma requires three-dimensional kinetic simulations that include pair creation due to gamma--gamma collisions, a computational challenge for the future.

We find that the presence of higher multipoles can significantly modify the geometry of the polar caps (defined as the angular region of current flow near the star).  In particular, one or both polar caps can be shifted away from the rotational pole, occupying a thin annulus at some finite angle $\theta_0$.  The width of the annulus scales with the rotation rate as $\epsilon$, in contrast to the $\sqrt{\epsilon}$ scaling of an ordinary polar cap.  Should these features persist in the case of an inclined rotator, they could potentially be observed if emission occurs close enough to the star for multipolar fields to be important.  In ordinary pulsars, the radio emission is believed to be produced at distances $R\sim20-50R_\star$, making the features unobservable.  However, in millisecond pulsars the emission likely originates within a few stellar radii, where multipolar fields can potentially survive \cite{Kramer98}.  Thus, we expect that observational signatures of shifted polar caps may be present for millisecond pulsars, but not ordinary pulsars.

One potential signature is double-peaked emission.  If pulsed emission from an inclined rotator mimics the shape of its annular polar cap, then the profile would be strongly double-peaked as each edge of the hollow beam sweeps through the telescope.  Although the details of the generation of pulsed emission are poorly understood, one might, in general, expect a more complicated beam shape when emission occurs near displaced polar caps, and hence more complicated beam shapes for millisecond pulsars.  There is general consensus that millisecond pulsars do indeed display more complicated beam shapes than ordinary pulsars (at least marginally \cite{Kramer98}), but this is difficult to quantify.  The next generation of radio telescopes (e.g., SKA \cite{Karastergiou15}) will significantly increase the sample of the observed millisecond pulsars and enable detailed studies of their emission geometry.

Another potential signature is a narrower pulse profile.  Shifted polar caps are parametrically narrower ($\sim\epsilon$ instead of $\sim\sqrt{\epsilon}$) than caps at the poles.  If only one of the two potential peaks is actually observed (for example, due to asymmetric synchrotron absorption of radio waves in the magnetosphere \cite{Beskin12}), then the main signature of shifted polar caps is a modified beam width scaling with rotation rate.  Tantalizingly, such a beam size scaling discrepancy has already been observed: millisecond pulsars are found to have systematically smaller beam widths compared to the usual scaling $\propto\sqrt{\epsilon}$ of normal radio pulsars \cite{Kramer98}.

A third potential signature is the phase-lag between gamma-ray and radio pulses.  Recent studies suggest that current sheets in the magnetosphere are natural sources of gamma rays \cite{Bai10, Uzdensky14, Cerutti16}.  Since the position of the current sheet is essentially unchanged, we expect gamma-ray light curves of multipolar magnetospheres to be identical to the dipolar case.  However, the radio beam shape may be significantly modified, as discussed above.  This could have important consequences for observations of radio to gamma-phase lags in millisecond pulsars \cite{Abdo10}.

Finally, shifted polar caps may have consequences for the pulsar wind by significantly enhancing the pair injection rate.  On a shifted polar cap, the curvature radius of the magnetic field is generally much shorter compared to a dipolar field close to the pole, rendering curvature radiation more efficient \cite{Barnard82}.  We have shown that spacelike current generically appears, and in particular covers a large portion of the polar cap in the quadrudipole model that we study in detail (Fig.~\ref{fig:Quadrudipole}).  These two facts favor efficient particle acceleration and emission of energetic curvature photons to produce secondary pairs.

\section*{Acknowledgements}

We thank Beno{\^i}t Cerutti, Konstantinos Gourgouliatos and Anatoly Spitkovsky for fruitful discussions and Alexander Tchekhovskoy for providing the simulation setup. This research was supported by the NASA Earth and Space Science Fellowship Program (grant NNX15AT50H to A.P.), as well as NSF grants 1205550 to Harvard University and 1506027 to the University of Arizona. The simulations presented in this paper used computational resources supported by the PICSciE-OIT High Performance Computing Center and Visualization Laboratory.

\appendix

\section{Split monopole}

The split monopole configuration can be solved analytically by perturbing the exact solution in the Schwarzschild spacetime \cite{Lyutikov11, Brennan13}.  This is the analog of the procedure used by Blandford \& Znajek \cite{Blandford77} to derive the split monopole solution for a slowly-spinning black hole.  Since the Kerr metric and the rotating star metric agree to linear order in $\epsilon$, the only difference is the boundary condition at the compact object.  In the black hole case, one finds $\Omega_F=\tfrac{1}{2}\Omega_H$, whereas here, we have $\Omega_F=\Omega$. This yields
\begin{subequations}
\label{eq:MichelMonopole}
\begin{align}
	\psi(r,\theta)&=\psi_o\pa{1-\ab{\cos{\theta}}},\\
	\label{eq:MichelCurrent}
	I(\psi)&=\pm2\pi\Omega\psi\pa{2-\frac{\psi}{\psi_o}}=\pm2\pi\Omega\sin^2{\theta},
\end{align}
\end{subequations}
where we take the $+$ sign (respectively, $-$ sign) for the northern (respectively, southern) hemisphere.

The last open field line lies on the equator,
\begin{align}
	\psi_o=\psi(r,\theta)\big|_{\theta=\pi/2}=B_0 R_\star^2.
\end{align}
The power is
\begin{align}
	\mathcal{L}=\frac{8\pi}{3}\psi_o^2\Omega^2.
\end{align}
The four-current is given by
\begin{subequations}
\label{eq:MichelMonopoleCurrent}
\begin{align}
	J^t&=\frac{2\psi_o}{\pa{r\alpha}^2}\pa{\Omega-\Omega_Z}\ab{\cos{\theta}},\\
	J^r&=-\frac{2\psi_o}{r^2}\Omega\ab{\cos{\theta}},\\
	J^\theta&=0,\\
	J^\phi&=0.
\end{align}
\end{subequations}
This agrees with the near-zone current expression Eq.~\eqref{eq:Current} using Eq.\eqref{eq:MichelMonopole}, but Eqs.~\eqref{eq:MichelMonopoleCurrent} are valid everywhere.  The norm of the current is
\begin{subequations}
\begin{align}
	J^2&=\pa{\frac{2B_0}{\alpha}\frac{R_\star^2}{r^2}}^2\pa{2\Omega-\Omega_Z}\Omega_Z\cos{\theta}^2\\
	&=\frac{B_0^2}{R_\star^2}h(\I,\C,\theta)\epsilon^2,
\end{align}
\end{subequations}
with
\begin{align}
	h=4\I\C\pa{\frac{2-\I\C}{1-\C}}\cos{\theta}^2.
\end{align}
We see that the current is spacelike everywhere for a reasonable star.  When evaluated on the pole, this agrees with the dipole results in Eqs.~\eqref{eq:NorthDipoleCurrentNorm} and \eqref{eq:NorthPoleDipoleCurrentNorm}, noting that $2B_1$ is the value of the magnetic field on the pole.

\begin{widetext}

\section{Solutions of the near-zone equation}
\label{app:Harmonics}

The solutions of Eq.~\eqref{eq:NearStreamEquation} are presented in \cite{Beskin10, Gralla16}.  The angular harmonics $\Theta_\ell$ are proportional to the Gegenbauer polynomials $\mathcal{G}_n^{(m)}$,
\begin{subequations}
\begin{numcases}
	{\Theta_\ell(\theta)=}
		{_2F_1}\!\br{\frac{\ell}{2},-\frac{\ell+1}{2};\frac{1}{2};\cos^2{\theta}}=-\frac{\Gamma\!\pa{-\frac{\ell}{2}}\Gamma\!\pa{\frac{\ell+1}{2}}}{2\sqrt{\pi}}\sin^2{\theta}\,\mathcal{G}^{(3/2)}_{\ell-1}(\cos{\theta})
		&$\ell$ odd,\\
		{_2F_1}\!\br{-\frac{\ell}{2},\frac{\ell+1}{2};\frac{3}{2};\cos^2{\theta}}\cos{\theta}=-(-1)^{\ell/2}\frac{\sqrt{\pi}\,\Gamma\!\pa{\frac{\ell}{2}}}{4\,\Gamma\!\pa{\frac{\ell+3}{2}}}\sin^2{\theta}\,\mathcal{G}^{(3/2)}_{\ell-1}(\cos{\theta})
		&$\ell$ even.
\end{numcases}
\end{subequations}
The first few harmonics are
\begin{align}
	\Theta_1(\theta)&=\sin^2{\theta},\\
	\Theta_2(\theta)&=\cos{\theta}\sin^2{\theta},\\
	\Theta_3(\theta)&=\pa{1-5\cos^2{\theta}}\sin^2{\theta},\\
	\Theta_4(\theta)&=\pa{1-\tfrac{7}{3}\cos^2{\theta}}\cos{\theta}\sin^2{\theta}.
\end{align}
We denote the radial harmonics regular at infinity by $R^>_\ell$.  These are given by
\begin{align}
	R^>_\ell(r)=-\frac{2}{\sqrt{\pi}}\pa{\frac{r}{4}}^{-\ell}\frac{\Gamma\!\pa{\ell+\frac{3}{2}}}{\pa{\ell+1}\Gamma(\ell)}\cu{{_2F_1}\!\br{\ell+2,\ell;1;1-\frac{2M}{r}}\log\pa{1-\frac{2M}{r}}+P_\ell\!\pa{\frac{r}{2M}}},
\end{align}
where the $P_\ell$ are polynomials recursively defined for $\ell=1,2,3,\ldots$ by
\begin{subequations}
\begin{align}
	P_1(x)&=x^2+\frac{x}{2},\qquad P_2(x)=4x^4-x^3-\frac{x^2}{6},\\
	P_\ell(x)&=\frac{\pa{2\ell-1}\br{\ell\pa{\ell-1}\pa{2x-1}-1}xP_{\ell-1}(x)-\ell^2\pa{\ell-2}x^2P_{\ell-2}(x)}{\pa{\ell+1}\pa{\ell-1}^2},
\end{align}
\end{subequations}
The normalization of the radial harmonics $R^>_\ell$ is chosen so that as $r\to\infty$, we have
\begin{align}
	R^>_\ell(r)=\frac{1}{r^\ell}+\mathcal{O}\!\pa{\frac{1}{r^{\ell+1}}}.
\end{align}
The first few harmonics are
\begin{subequations}
\begin{align}
	R^>_1(r)&=-\frac{3}{2r}\br{\frac{3-4\alpha^2+\alpha^4+4\log{\alpha}}{\pa{1-\alpha^2}^3}}, \label{eq:R1} \\
	R^>_2(r)&=-\frac{10}{3r^2}\br{\frac{17-9\alpha^2-9\alpha^4+\alpha^6+12\pa{1+3\alpha^2}\log{\alpha}}{\pa{1-\alpha^2}^5}}, \label{eq:R2}\\
	R^>_3(r)&=-\frac{35}{4r^3}\br{\frac{43+80\alpha^2-108\alpha^4-16\alpha^6+\alpha^8+24\pa{1+8\alpha^2+6\alpha^4}\log{\alpha}}{\pa{1-\alpha^2}^7}},\label{R3}
\end{align}
\end{subequations}
where $\alpha=\sqrt{1-2M/r}$ was introduced in Eq.~\eqref{eq:AlphaOmegaDefinition}.

\section{Quadrudipole formulas}
\label{app:Quadrudipole}

The exact solution of Eq.~\eqref{eq:QuadrudipolePolarCaps} is 
\begin{align}
	\label{eq:ExactCriticalAngles}
	\cos\theta_\pm=\frac{1}{3q}\br{\frac{1\pm i\sqrt{3}}{2^{2/3}}\frac{1+3q^2}{X}+\frac{1\mp i\sqrt{3}}{2^{4/3}}X-1},
\end{align}
where we introduced
\begin{subequations}
\begin{align}
	X^3&\equiv Y+\sqrt{Y^2-4\pa{1+3q^2}^3},\\
	Y&\equiv 2+9\pa{3\frac{n}{\Delta_1}\epsilon-2}q^2.
\end{align}
\end{subequations}
For small $\epsilon$, we can approximate this as
\begin{align}
	\label{eq:ApproximateCriticalAngles}
	\theta_+=\sqrt{\frac{n\epsilon}{\pa{1+q}\Delta_1}},\qquad\theta_-=
	\begin{cases} 
		\pi-\sqrt{\frac{n\epsilon}{\pa{1-q}\Delta_1}}&q<1,\\
		\pi-\sqrt{q-1+\sqrt{\frac{2n\epsilon}{\Delta_1}+\pa{q-1}^2}},&q\sim1,\\
		\theta_0-\frac{q^2}{\pa{q^2-1}^{3/2}}\frac{n\epsilon}{\Delta_1},&q>1.
	\end{cases}
\end{align}
The first and third lines correspond to Eqs.~\eqref{eq:SmallQuadrudipole} and \eqref{eq:LargeQuadrudipole}; now, we add an expression for the transition region.  Equation~\eqref{eq:ApproximateCriticalAngles} is convenient for seeing the scaling with $\epsilon$ in the various regimes, but for studying the problem globally it is generally easier to work with the single expression \eqref{eq:ExactCriticalAngles}.  We can view Eq.~\eqref{eq:ExactCriticalAngles} as one convenient way to smoothly transition between the regimes by adding in higher-order terms in $\epsilon$.

In terms of dimensionless coordinates $\delta=r/R_\star$ and $x=\sin{\theta}$, the quadrudipole current norm $J^2$ is given everywhere by
\begin{align}
	\label{eq:QuadrudipoleCurrentNorm}
	J^2&=\pa{\frac{B_1}{R_\star}}^2\frac{\pa{2\epsilon}^2}{\pa{c-\delta}\delta^3\Lambda_1}g(\I,\C,Q,\delta,x),
\end{align}
where
\begin{subequations}
\begin{align}
	g&=\br{\frac{A^2x^2}{\delta\pa{\C-\delta}}+\frac{\br{\pa{3\Sigma-\Lambda_1}x^2-2\Sigma}^2}{1-x^2}}\br{\frac{2}{5}\pa{\frac{\Sigma x^2}{n\epsilon}}^3+\frac{\Sigma x^2}{n\epsilon}-1}^2-\br{\frac{Bx^2}{\delta\pa{\C-\delta}}-2\Sigma}^2\pa{1-\frac{\I\C}{\delta^3}}^2,\\
	\Sigma&=\Lambda_1+\pa{Q-1}\Lambda_2=\Lambda_1\pa{1+q\cos{\theta}},\qquad Q=1+\frac{\Lambda_1}{\Lambda_2} q \cos{\theta},\\
	A&=3+2\pa{\C-\delta}\Lambda_1-\frac{20}{\C^2}\pa{Q-1}\br{3\C-4\delta+2\pa{\C-2\delta}\pa{\C-\delta}},\\
	B&=\frac{9}{2}\C-3\delta+3\C\pa{\C-\delta}\Lambda_1-10\pa{Q-1}\br{9-10\frac{\delta}{\C}+2\pa{3-5\frac{\delta}{\C}}\pa{\C-\delta}\Lambda_1}.
\end{align}
\end{subequations}
Here, $\Lambda_{1,2}$ are dimensionless radial harmonics $R^>_{1,2}$ scaled to reduce to $\Delta_{1,2}$ at the star,
\begin{align}
	\Lambda_1(\delta)=-\frac{3}{2\C^3}\br{\C\pa{\C+2\delta}+2\delta^2\log\pa{1-\tfrac{\C}{\delta}}},\qquad\Lambda_2(\delta)=-\frac{20}{3\C^2}\br{4+\pa{3\C-4\delta}\Lambda_1(\delta)},\qquad\Lambda_{1,2}(1)=\Delta_{1,2}.
\end{align}
For $q>1$, when evaluated at $\theta_0=\pi-\arctan\sqrt{q^2-1}$ (corresponding to $\Sigma=0$), the result is
\begin{subequations}
\label{eq:QuadrudipoleCurrentNormPolarCap}
\begin{align}
	J^2|_{\theta=\theta_0} &=\pa{\frac{B_1}{R_\star}}^2\frac{\pa{2q\epsilon}^2}{1-\C}\pa{1-\frac{1}{q^2}}^2\cu{1+\pa{\frac{20}{\Delta_1\Delta_2\C^2}}^2\br{\frac{U^2}{\pa{1-\C}\pa{q^2-1}}-\pa{\frac{V}{q}}^2\pa{\frac{1-\I\C}{1-\C}}^2}},\\
	U&=\frac{4}{3}\Delta_1\pa{\Delta_1+2}\pa{1-\C}-4,\qquad V=\Delta_1^2\pa{1-\C}\C-4\Delta_1\pa{1-\C}^2+2\pa{2-3\C}.
\end{align}
\end{subequations}
\hfill\includegraphics[scale=0.03]{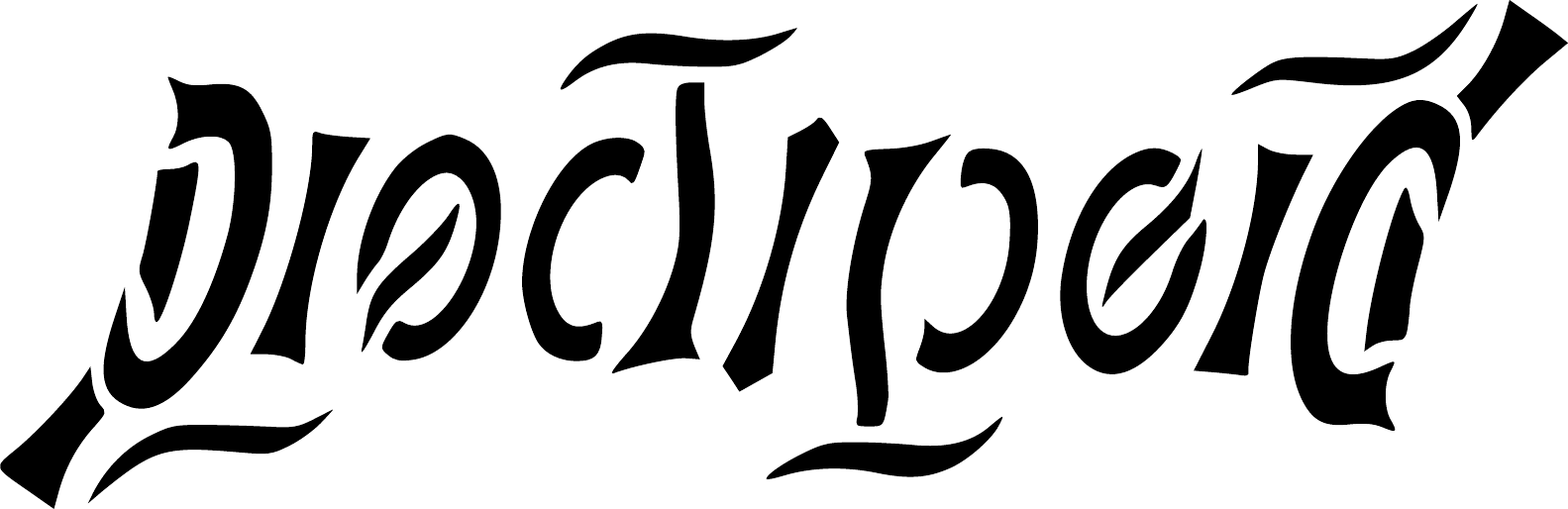}

\clearpage

\end{widetext}

\bibliographystyle{apsrev4-1}
\bibliography{PulsarMagnetospheres}

\end{document}